\documentclass[twocolumn,  
 nofootinbib,   
 showpacs,  
groupedaddress,  preprintnumbers,  aps,  prd]{revtex4-1}
\usepackage{amsmath,  amssymb,   amsthm,  mathrsfs}
\usepackage{color}
\usepackage[subnum]{cases}

\newcommand{\be}{\begin{equation}}
\newcommand{\bea}{\begin{eqnarray}}
\newcommand{\eea}{\end{eqnarray}}
\newcommand{\ba}{\begin{align}}
\newcommand{\ea}{\end{align}}
\newcommand{\ee}{\end{equation}}

\newcommand{\p}{\partial}

\newcommand{\Tr}{\mathrm{Tr}\,}

\begin{document}
\preprint{}
\title{Holographic Fidelity Susceptibility }

\author{Mohsen Alishahiha$^a$}
\author{Amin Faraji Astaneh$^b$}

\affiliation{$^a$School of Physics, $^b$School of Particles and Accelerators,  Institute for Research in Fundamental Sciences (IPM)\\
P.O. Box 19395-5531,   Tehran,   Iran\\ 
{\rm email:  alishah@ipm.ir, faraji@ipm.ir}}

\begin{abstract}
For a field theory with a gravitational dual,  we study holographic  fidelity susceptibility for two 
states related by a deformation of a relevant  operator. To do so, we study back reaction of a
massive scalar field on the  asymptotic 
behavior of the  metric in an Einstein gravity coupled to a  
massive scalar field. Identifying the two states, holographically,  
with the  original and back reacted geometries, the corresponding holographic fidelity 
susceptibility is given by the difference  of the volume of an extremal time slice 
evaluated on the original and  back  reacted geometries.
\end{abstract}


\maketitle

\section{Introduction}\label{intro}

Recent progress on black hole physics has opened up a possibility to make a connection 
between quantum information theory and back hole physics (see \cite{Almheiri:2012rt}
and its citations).  Most of our understanding on this subject is due to the 
 gauge/gravity duality. In particular    holographic entanglement entropy \cite{Ryu:2006bv} 
may be thought of as an example which  has made this connection more concrete.   
Computational complexity is another example\cite{Susskind:2014rva}.

More recently, motivated by complexity/volume duality it was proposed that 
the volume of a co-dimension one time slice may be related to the fidelity appearing 
in the literature of  quantum information theory\cite{MIyaji:2015mia}. 
 Fidelity  is a quantity that measures 
similarity or  distance between two states\cite{Uhlmann:1976}. 
When  both states under consideration  are pure it is given by the inner 
product of two states. To be more precise,  let us start with a quantum pure state 
$|\Psi(\lambda)\rangle$ in the Hilbert of a quantum system with $\lambda$ being  a tunable 
parameter of the model. Then for two pure states  parametrized by $\lambda_1$ and $\lambda_2$ 
the fidelity is
\be
{\cal F}=|\langle \Psi(\lambda_2)|\Psi(\lambda_1)\rangle |,
\ee
which for small $\delta \lambda=\lambda_2-\lambda_1$ it can be expanded as follows
\be
{\cal F}=1- G_\lambda (\delta\lambda)^2+{\cal O }(\delta\lambda^3),
\ee
where $G_\lambda$ is fidelity susceptibility (see \cite{Gu:2008} for a review). 
If one perturbes the first state by an operator with dimension $\Delta$, up to a constant 
of order of one factor, the corresponding fidelity susceptibility will be scaled as (see {\it e.g.}
  \cite{MIyaji:2015mia})
  \be\label{FS}
G_\Delta\sim \frac{L^{2\Delta-d-2}}{\epsilon^{2\Delta-d-2}},
\ee
where $\epsilon $ is a UV cut off and $L$ is a length scale of the model.

It is proposed  that when the system is deformed by an exact marginal operator,
$\Delta=d+1$,  
the  holographic dual of the fidelity susceptibility is given by the volume of a
certain  co-dimension one hypersurface in an AdS geometry
(see \cite{Alishahiha:2015rta} for reduced fidelity). 

The aim of this letter is to make this correspondence more precise and in particular 
to extend it for a case where the system is deformed by a relevant operator. 
Holographically, we will do that
by studying  the back reaction  of a massive scalar field on the metric 
in an Einstein gravity coupled to a massive scaler field.    
We should admit that most of the materials we need have been already presented rather 
extensively in the literature as we shall review in section two. The main contribution of the
present letter is to come up with a new application of the results given in section three. 


\section{Review of background geometry} 

In this section we shall present the asymptotic behaviors of the metric and a scalar field 
for a gravitational theory whose action is given by 
\be\label{action1}
I\!=\!-\frac{1}{16\pi G_N}\int \!d^{d+2}x\sqrt{-G}\left(\!R-\frac{1}{2}(\p\Phi)^2-V(\Phi)\!\right)\, ,
\ee
where $V(\Phi)=-\frac{d(d+1)}{L^2}+\frac{1}{2}m^2\Phi^2$ and $L$ is a length scale 
of the model.  An asymptotic AdS solution of the model
taking into account the back reaction of the scalar field may be given by \cite{Henningson:1998gx}
\bea
ds^2\!&\!=&\!G_{\mu\nu}dX^\mu dX^\nu=\frac{L^2}{4\rho^2}d\rho^2+
\frac{1}{\rho}g_{ij}(x^i,\rho)dx^idx^j\, ,\cr
&&\Phi(x,\rho)=\rho^{\alpha/2}(\phi_{0}(x^i)+\rho\, \phi_{1}(x^i)+\cdots)\, .
\eea
with $i,j=0,1\cdots,d$ and
\bea
g_{ij}(x^i,\rho)\!&\!=\!&\! g_{ij}^{(0)}(x^i)+\rho\, g_{ij}^{(1)}(x^i)+\cdots \cr &&
+\rho^\alpha\left(g_{ij}^{(\alpha)}(x^i)+\rho\, 
g_{ij}^{(\alpha+1)}(x^i)+\cdots\right)  \,.
\eea
Here $\alpha={(d+1)}/{2}-\sqrt{{(d+1)^2}/{4}+m^2L^2}$ is related to the dimension of an
operator which is dual to the scalar field; $\Delta=d+1-\alpha$. We shall consider relevant 
operator; $0\leq \alpha<(d+1)/2$. Note that the case of $\alpha=(d+1)/2$ is 
excluded due to that fact that in this case  $g_{ij}^\alpha$ cannot be fixed 
by perturbative expansion  around $\rho=0$.  Solving  the corresponding 
equations of motion order by order in power of $\rho$ near the boundary ($\rho\rightarrow 0$)
one may find all needed functions appearing in 
the above expressions in terms of the sources  $g_{ij}^{(0)}(x^i)$ and $\phi_{0}(x^i)$
(for $d>1$) \cite{Hung:2011ta}
 \be\label{ansatz}
\begin{split}
&g^{(1)}_{ij}=-\frac{L^2}{d-1}\left( R^{(0)}_{ij}-\frac{1}{2d} R^{(0)}g^{(0)}_{ij}\right),\,\,\,\,
g^{(\alpha)}_{ij}=-\frac{\phi_{0}^2}{4d}g^{(0)}_{ij}\, ,\\
&g^{(\alpha+1)}_{ij}=c_1\phi_{0}^2R^{(0)}_{ij}+c_2\phi_{0}^2R^{(0)}g^{(0)}_{ij} +c_3
\p_i\phi_{0}\p_j\phi_{0}\\
&+c_4\nabla_i\p_j\phi_{0}^2+c_5 g^{(0)}_{ij}\Box \phi_{0}^2+c_6 g^{(0)}_{ij}
(\p\phi_{0})^2\, ,
\end{split}
\ee
where 
\bea
&&c_1=- \alpha\frac{b(\alpha+2)+2(\alpha+1)a}{(d-1)(\alpha+1)}\, \;\;\;
c_2=-\frac{1}{2d}(c_1+2\alpha c_5)\, ,\cr
&&c_3=-\frac{2\alpha+3}{\alpha+1}b-\frac{2(2\alpha+1)}{\alpha}a
\,\;\;\;c_4=\frac{b}{2}+a\, ,\\
&&c_5=-\frac{2\alpha d }{(\alpha+1)(d-2\alpha-1)}\, , \, c_6=\frac{4(\alpha-1)d}{(\alpha+1)
(d-2\alpha-1)}\, .\nonumber
\eea
with $a=\alpha\, c_6L^2-c_5(d-4\alpha-1)L^2$ and $b=-\frac{L^2}{4d}$. Moreover one finds
\cite{Hung:2011ta}
\be\label{scalar}
\phi_{1}=\frac{L^2}{2(d-2\alpha-1)}\left( \Box\phi_{0}-\frac{1}{2d} R^{(0)}
\phi_{0
}\right)\, .
\ee
Here for simplicity we have dropped the explicit dependence of  functions on $x^i$.
Note also that the term corresponding to $c_6$ was missed in 
\cite{Hung:2011ta}, see however\cite{Mir:2013pca}. To the best of our knowledge 
no body has explicitly computed $c_5\; {\rm and}\; c_6$, previously.



 \section{Change of volume and fidelity }
In this section we would like to study the leading divergent term of the volume of  an extremal
co-dimension  one hypersurface in the back reacted geometry presented   in the 
previous section.  To proceed, consider a time slice  on the boundary whose extension to the bulk 
is the desired extremal hypersurface. The profile of the corresponding  hypersurface in the bulk
is given by $X^{\mu}(Y^A)$ with $Y^A=(y^a,\sigma)$, and  $a=1,\cdots, d$ are coordinates of 
it. The induced metric on the hypersurface is 
\be\label{h1}
h_{AB}[Y^A]=\p_A X^\mu\p_B X^\nu G_{\mu\nu}[X].
\ee
Following \cite{{MIyaji:2015mia}, {Alishahiha:2015rta}} one is interested in 
the volume of the corresponding hypersurface (see also \cite{Carmi:2016wjl})
\be
{\cal C}=\frac{1}{8\pi G_N L}\int d^{d+1}Y\,\sqrt{\det h_{AB}}.
\ee
Extremizing the volume one arrives at 
\be\label{CC}
\p^A\p_A X^\mu+\Gamma^\mu_{\alpha\beta}\p^AX^\alpha\p_AX^\beta-h^{AB}
\gamma^C_{AB}\p_CX^\mu=0,
\ee
that is indeed the trace of the \emph{Gauss-Weingarten} equation. It might be thought of 
as equations of motion for $X^\mu$.  Here 
$\Gamma^\mu_{\alpha\beta}$ and $\gamma^C_{AB}$ are connections constructed from 
metrics $G_{\mu\nu}$ and 
$h_{AB}$, respectively.

It is useful to take a gauge in which $\sigma=\rho$ and $h_{a\rho}=0$. In this  
gauge one may consider the following expansion for the profile of the extremal surface
\cite{Graham:1999pm}
\be
X^i(\rho,y^a)=X^{(0)i}(y^a)+\rho\, X^{(1)i}(y^a)+\cdots\, .
\ee
From the equation \eqref{CC} one has 
\be\label{h2}
X^{(1)i}=\frac{L^2K }{2d}n^i\, ,
\ee
where $K$ is the trace of the extrinsic curvature of the time slice. By making use of 
equations \eqref{h1} and \eqref{h2} it is straightforward to find the asymptotic 
expansion of the induced metric as follows
\bea\label{IM}
h_{\rho\rho}&=&\frac{L^2}{4\rho^2}(1+\rho h^{(1)}_{\rho\rho}+\rho^{\alpha+1} h^{(\alpha+1)}_{\rho\rho})+\cdots\, ,\\
h_{ab}&=&\frac{1}{\rho}\left[h^{(0)}_{ab}+\rho h^{(1)}_{ab}+\rho^\alpha  (h^{(\alpha)}_{ab}
+\rho h^{(\alpha+1)}_{ab})\right]+\cdots\, ,\nonumber
\eea
where $h_{ab}^{(\alpha)}= g^{(\alpha)}_{ab}$ and
\bea
\!&&\!h^{(1)}_{\rho\rho}
=-\frac{L^2K^2}{d^2},\;\;\;\;\;\;\;\;\;h^{(1)}_{ab}=g^{(1)}_{ab}+\frac{L^2K}{d}K_{ab}\,\\
\!&&\!h^{(\alpha+1)}_{\rho\rho}=\frac{L^2K^2\phi_{0}^2}{4d^3},
\;\;\; h^{(\alpha+1)}_{ab}=g^{(\alpha+1)}_{ab}-\frac{L^2\phi_0^2K}{4d^2}K_{ab}.
\nonumber
\eea
Here $g^{(n)}_{ab}=e^i_ae^j_bg^{(n)}_{ij}$ is the projection of the boundary metric
into the time slice and  $e^i_a=\p_aX^j$. Note that all functions are functions of 
$y_a$'s. Note also that $h^{(1)}_{\rho\rho}$ and $h^{(1)}_{ab}$ have been already computed 
in \cite{Carmi:2016wjl}.

Using the asymptotic expression for the induced metric \eqref{IM} one finds
\bea
&&\delta{\cal C}=\frac{1}{8\pi G_N}\int d^dy\,d\rho \frac{\sqrt{h^{(0)}}}
{4\rho^{\frac{d+2}{2}-\alpha}}\bigg[\Tr h^{(\alpha)}_{ab}\\&&
\;\;\;\;\;\;\;\;+\rho\bigg( 
h^{(\alpha+1)}_{\rho\rho}
+\Tr h^{(\alpha+1)}_{ab}+\Tr (h^{(1)} h^{(\alpha)})_{ab}\cr &&
\;\;\;\;\;\;\;\;+\frac{1}{2} \Tr h^{(1)}_{ab} \Tr h^{(\alpha)}_{ab}+\frac{1}{2}
h^{(1)}_{\rho\rho}\Tr h_{ab}^{(\alpha)}
\bigg)\!\bigg]+\cdots\, .\nonumber
\eea
Thus, setting a regulator surface at $\rho=\epsilon^2/L^2$,  at  the leading order  the 
change of the volume   due a relevant operator with dimension $\Delta$ is 
\bea\label{dC}
&&\delta {\cal C}_\Delta\!=\!-\frac{1}{64(2\Delta-d-2)G_N}\,\frac{L^{2\Delta-d-2}}
{\epsilon^{2\Delta-d-2}}\!
\int\! d^{d}y\sqrt{h^{(0)}}\;\bigg[\phi_{0}^2\, \cr
 &&\;\;\;\;\;\;\;\;\;\;+
\frac{\epsilon^2}{L^2}\bigg(\Tr g^{(\alpha+1)}_{ab}-
\frac{\phi_0^2}{8d}(d+2)\Tr g^{(1)}_{ab}\cr
&&\;\;\;\;\;\;\;\;\;\;\;\;\;\;\;\;\;\;\;\;\;\;-\frac{L^2K^2\phi_0^2}{8d^3}
(d^2+3d-2)\bigg)+\cdots\bigg].
\eea
For flat time slice and constant $\phi_0$ the above expression reads
\be
\delta {\cal C}_\Delta=-\frac{\phi_0^2V_d}{64(2\Delta-d-2)G_N}\,\frac{L^{2\Delta-d-2}}
{\epsilon^{2\Delta-d-2}}
\ee
where $V_d$ is the volume of time slice on the boundary. This should be compared with equation \eqref{FS}.

It is worth noting that for a deformation by an operator with dimension $\Delta=
\frac{d}{2}+1$ one gets logarithmic divergence as follows
\be
\delta {\cal C}_\Delta=-\frac{\int d^{d}y\sqrt{h^{(0)}}
\phi_{0}^2}{64dG_N}\,\log\frac{L}{\epsilon}  \, .
\ee

To conclude  one may propose that the change of volume of the extremal co-dimension 
one hypersurface  due to the effect of  a relevant operator could provide a holographic 
description for fidelity susceptibility in the dual field theory. 
It should be noted that the logarithmic 
behavior for fidelity susceptibility has been also found in the literature \cite{Gu:2008}.

To further justify  this proposal it is constructive  to extend 
our discussions for the case where the model has a non-trivial dynamical scaling. Indeed
fidelity susceptibility for a deformation caused by an operator with dimension $\Delta$ in a 
field theory with a dynamical exponent  scales as  
$G_\Delta\sim \epsilon^{-2\Delta+d+2z}$,
where $z$ is the dynamical exponent\cite{Gu:2008}. It is our aim 
 to get this  expression using a holographic model. To proceed, we note that holographic description of field theories 
with non-trivial exponent  may be provided by  Lifshitz geometries\cite{Kachru:2008yh}. 
\be
dS^2=-\frac{dt^2}{r^{2z}}+\frac{dr^2}{r^2}+\frac{\sum_{i=1}^ddx_i^2}{r^2}.
\ee
To study  holographic fidelity for a deformation caused by a relevant operator in this mode, one
may consider a massive scalar field in the above geometry.  Using the corresponding equations of motion one finds  the asymptotic behavior of the scalar field as follows 
\be 
\Phi(x,r)=r^\xi(\phi_0+r^2\phi_1+\cdots),
\ee
where $\xi=(d+z)/2-\sqrt{(d+z)^2/4+m^2}$ and the dimension of the corresponding dual
operator is $\Delta=d+z-\xi$. To find the change of the volume one needs to 
 study back reactions of this scalar field on the metric. To do so, essentially we should go 
 through the procedure we have presented in the previous section for 
 an asymptotic AdS solution. Let us just present the arguments and the results 
 and postpone the details to an extend version of this paper. Actually  
  from the asymptotic form of  the 
 scalar field one finds  that  the corresponding corrections of the back reacted metric 
start at the order of  $\delta g_{ij}\sim r^{2\xi} g_{ij}^{(\xi)}$. 
Therefore the leading correction to the induced metric of the extremal hypersurface is
 $\delta h_{ab}\sim r^{2\xi}h_{ab}^{(\xi)}$.
It is then straightforward to find the leading correction to the  change of  the volume of
the extremal hypersurface associated to the time slice $t$=constant as follows
\be
\delta {\cal C}_\Delta\!\sim\! \int _\epsilon\! drd^dy \frac{\sqrt{h^{(0)}}}{r^{d+1}}r^{2\xi}
\Tr h^{(\xi)}_{ab}
\sim\frac{\int \! d^dy \sqrt{h^{(0)}}\phi_0^2}{\epsilon^{2\Delta-d-2z}} ,
\ee
in agreement with the field theory result \cite{Gu:2008}. In particular for a marginal operator, $\Delta=
d+z$,  one gets $\epsilon^{-d}$ that is independent of $z$, as expected.

\section{Discussions}

In this paper we have studied holographic fidelity susceptibility associated with a relevant 
operator with dimension $\Delta$ satisfying $(d+1)/2<\Delta\leq d+1$. Holographically 
the corresponding fidelity susceptibility is computed by the change of the volume of an
extremal co-dimension one time slice  due to the change of the background geometry.  

Actually the proposal makes a rough connection between holographic boundary state
and the volume of an extremal time slice in the bulk (holographic complexity) as 
$|\psi(\lambda)\rangle =e^{-{\cal C}_\lambda}$ ($\langle \psi(\lambda) |=e^{{\cal C}_\lambda}$). 
Thus one has
\be
\langle \psi(\lambda_2)|\psi(\lambda_1)\rangle=e^{\delta {\cal C}_\Delta}\sim 1+\delta {\cal C}_\Delta.
\ee
Here we have assumed that two states are related by a relevant operator with dimension $\Delta$.
Note also that, as we have seen, $\delta {\cal C}_\Delta<0$.

It is worth mentioning that although our holographic approach applies for relevant operators with 
the range of dimensions given above, for $(d+1)/2<\Delta<(d+2)/2$ the resultant fidelity 
susceptibility is not  UV divergent. For  particular value of  $\Delta=(d+2)/2$ the most divergent term 
is a logarithmic, in agreement with the field theory result\cite{Gu:2008}.

We have found that the main contribution to the most divergent term is given by 
$\rho^{\alpha}\Tr g^{(\alpha)}_{ab}$ term  with $0\leq \alpha\leq d/2$. The lower bound 
corresponds to the marginal operator and the upper bound to that of logarithmic divergent.
 Note that in this range our procedure also works for $d=1$ case. 

More generally for a $d+1$ dimensional field theory  with a dynamical exponent $ z$ we 
have found that holographic fidelity susceptibility is scaled as follows
\be\label{SAUB}
G_\Delta\sim\Bigg\{ \begin{array}{rcl}
&(\frac{L}{\epsilon})^{2\Delta-d-2z}&\,\,\,2\Delta-2z\neq d,\\
&\log \frac{L}{\epsilon}&\,\,\,2\Delta-2z=d.
\end{array}
\ee
In this letter our main focus was on the most divergent term in the holographic fidelity susceptibility
associated with a relevant operator. In general at subleading order one may also 
get logarithmic divergent term. In fact the logarithmic divergent term appears when $d$ is even. Moreover, it may also contain a finite term that our asymptotic  study cannot  capture 
it. Indeed to find the full expression one requires to have the full back reacted geometry.

So far we have considered cases where two states are related by a deformation caused by 
a relevant operator. One may also consider a case where the deformation is given by the change 
of the temperature. In this case it is known that the corresponding fidelity susceptibility is indeed
the heat capacity. It is then illustrative to study this case in our framework
(see also \cite{{Banerjee:2017qti},{Momeni:2017mmc}}).

Holographically, a thermal state may be associated to a black hole solution. Therefore to find 
the corresponding fidelity susceptibility one needs to compute the change  of the volume of
the corresponding extremal co-dimension one time slice when we change an  AdS geometry 
to an AdS black hole.
Doing so one finds \cite{MIyaji:2015mia} 
\be
\delta {\cal C}=b_d\frac{V_d L^dd}{4G \rho_H^d},
\ee
where $\rho_H$ is the radius of horizon and $b_d$ is a numerical factor (see Eq 25 of \cite{MIyaji:2015mia}). On the other hand computing the heat capacity of the corresponding black hole one gets (see {\it e.g.} \cite{Brown:1994gs})
\be
 C_v=\frac{V_d L^dd}{4G \rho_H^d},
\ee
that is equal to the holographic fidelity susceptibility up to an order of one numerical factor $b_d$.
It is, however, important to note that for $d=1$, one has $b_1=0$ which is not compatible with the 
above picture. Of course this can be understood since a BTZ black hole is locally AdS and therefore the change of  volume is zero.  Indeed AdS and BTZ black hole can be 
distinguished from their global structures. An Euclidean BTZ black hole has  a topology
of a torus and thus using  an SL(2) transformation one may compute the volume of a co-dimension 
one time slice after a Wick rotation by which the variation of the volume leads to $\delta{\cal C}=
\frac{V_1L}{4G\rho_H} $, in agreement with black hole result.

\begin{acknowledgments}

We would like to thank A. Akhavan, A. Mollabashi, M. R. Mohammadi Mozaffar,  A. Naseh, A. Shirzad,
 M. R.  Tanhayi and M. H. Vahidinia for useful discussions.  We would also like to thank T. Takayanagi
 for a correspondence.  A.F.A would like to thank the CERN TH-department for very kind hospitality during working on this project. This work is supported by Iranian National Science
 Fundation (INSF).

\end{acknowledgments}


\begin{thebibliography}{}







\bibitem{Almheiri:2012rt} 
  A.~Almheiri, D.~Marolf, J.~Polchinski and J.~Sully,
  ``Black Holes: Complementarity or Firewalls?,''
  JHEP {\bf 1302}, 062 (2013)
  [arXiv:1207.3123 [hep-th]].


\bibitem{Ryu:2006bv} 
  S.~Ryu and T.~Takayanagi,  
  ``Holographic derivation of entanglement entropy from AdS/CFT,  ''
  Phys.\ Rev.\ Lett.\  {\bf 96},   181602 (2006)
  [hep-th/0603001].


\bibitem{Susskind:2014rva} 
  L.~Susskind,  
  ``Computational Complexity and Black Hole Horizons,  ''
  arXiv:1402.5674 [hep-th].


\bibitem{MIyaji:2015mia} 
  M.~Miyaji,   T.~Numasawa,   N.~Shiba,   T.~Takayanagi and K.~Watanabe,  
  ``Gravity Dual of Quantum Information Metric,  ''
  arXiv:1507.07555 [hep-th].


\bibitem{Uhlmann:1976}
A.~ Uhlmann, `` The transition probability in the state space of A$^*$-algebra,''
Rep.\  Math.\  Phys.\  {\bf 9} 273 (1976), 


\bibitem{Gu:2008}
S-J.~ Gu,   ``Fidelity Approach to Quantum Phase Transi-
tions,''  Int.\ J.\ Mod.\ Phys.\ B {\bf 24} (2010) 4371 
[arXiv:0811.3127 [quant-ph]].


\bibitem{Alishahiha:2015rta} 
  M.~Alishahiha,
  ``Holographic Complexity,''
  Phys.\ Rev.\ D {\bf 92}, no. 12, 126009 (2015)
  [arXiv:1509.06614 [hep-th]].


 
\bibitem{Henningson:1998gx} 
  M.~Henningson and K.~Skenderis,
  ``The Holographic Weyl anomaly,''
  JHEP {\bf 9807}, 023 (1998)
  [hep-th/9806087].
 
 
 
\bibitem{Hung:2011ta} 
  L.~Y.~Hung, R.~C.~Myers and M.~Smolkin,
  ``Some Calculable Contributions to Holographic Entanglement Entropy,''
  JHEP {\bf 1108}, 039 (2011)
  [arXiv:1105.6055 [hep-th]].
 
 
\bibitem{Mir:2013pca} 
  M.~Mir,
  ``On Holographic Weyl Anomaly,''
  JHEP {\bf 1310}, 084 (2013)
 [arXiv:1307.5514 [hep-th]].
 
 
 
\bibitem{Carmi:2016wjl} 
  D.~Carmi, R.~C.~Myers and P.~Rath,
  ``Comments on Holographic Complexity,''
  JHEP {\bf 1703}, 118 (2017)
  [arXiv:1612.00433 [hep-th]].

 
\bibitem{Graham:1999pm} 
  C.~R.~Graham and E.~Witten,
  ``Conformal anomaly of submanifold observables in AdS / CFT correspondence,''
  Nucl.\ Phys.\ B {\bf 546}, 52 (1999)
  [hep-th/9901021].
 
 




\bibitem{Kachru:2008yh} 
  S.~Kachru, X.~Liu and M.~Mulligan,
  ``Gravity duals of Lifshitz-like fixed points,''
  Phys.\ Rev.\ D {\bf 78}, 106005 (2008)
  [arXiv:0808.1725 [hep-th]].

\bibitem{Banerjee:2017qti} 
  S.~Banerjee, J.~Erdmenger and D.~Sarkar,
  ``Connecting Fisher information to bulk entanglement in holography,''
  arXiv:1701.02319 [hep-th].
  
\bibitem{Momeni:2017mmc} 
  D.~Momeni, A.~Myrzakul, R.~Myrzakulov, S.~Alsaleh and L.~Alasfar,
  ``Thermodynamics of AdS spacetime via Regularized Fidelity Susceptibility,''
  arXiv:1704.05785 [hep-th].
  
\bibitem{Brown:1994gs} 
  J.~D.~Brown, J.~Creighton and R.~B.~Mann,
  ``Temperature, energy and heat capacity of asymptotically anti-de Sitter black holes,''
  Phys.\ Rev.\ D {\bf 50}, 6394 (1994)
  [gr-qc/9405007].
 
  
  
\end{thebibliography}
\end{document}